\documentclass[12pt,a4,fullpage,amssymbols]{article}
\usepackage{amssymb, amsmath}
\renewcommand{\theequation}{\arabic{section}.\arabic{equation}}

\def\btheta{\mbox{\boldmath $\theta$}}

\begin{document}
\title{An exact and explicit formula for pricing lookback options with regime switching}
\author{Leunglung Chan\footnote {School of Mathematics and Statistics, University of New South Wales, Sydney, NSW, 2052, Australia, Email: leung.chan@unsw.edu.au} and Song-Ping Zhu\footnote {School of Mathematics and Applied Statistics, University of Wollongong, Wollongong,
NSW 2522, Australia, Email: spz@uow.edu.au}}
\maketitle
\begin{abstract}
This paper investigates the pricing of European-style lookback options when the price dynamics of the underlying risky asset are assumed to follow a Markov-modulated Geometric Brownian motion; that is, the appreciation rate and the volatility of the underlying risky asset depend on unobservable states of the economy described by a continuous-time hidden Markov chain process. We derive an exact, explicit and closed-form solution for European-style lookback options in a two-state regime switching model.
\end{abstract}
\section{Introduction}
Option pricing is an important field of research in financial economics from both a theoretical and practical
point of view. The pioneering work of Black and Scholes (1973) and Merton (1973) laid the foundations
of the field and stimulated important research in option pricing theory, its mathematical models and its computational techniques.
The Black-Scholes-Merton formula has been widely adopted by traders,
analysts and investors. Despite its popularity the Black-Scholes-Merton formula has been documented in many
studies in empirical finance that the Geometric Brownian Motion (GBM) assumed in the Black-Scholes-Merton model does not provide a realistic
description for the behavior of asset price dynamics. During the past few decades many extensions to the Black-Scholes-Merton model have been
introduced in the literature to provide more realistic descriptions for asset price dynamics. In particular, many models have been introduced
to explain the empirical behavior of the implied volatility smile and smirk. Such models include the stochastic volatility models,
jump-diffusion models and models driven by L\'{e}vy process.

Recently, there has been considerable interest in applications of regime
switching models driven by a Markov chain to various financial
problems. For an overview of Markov chains, see Elliott et al. (1994). Guo (2001) investigated European option pricing
problem under a regime switching model. Buffington
and Elliott (2002a,b) considered the option pricing problems for
European and American options in a Black-Scholes market in
which the states of the economy are described by a finite state Markov process. Boyle and Draviam proposed a
numerical method to solve the system of coupled partial differential equations for the price of exotic options under regime switching.
Zhu et al. (2012) derived a closed-form solution for European options in a two-state regime switching model. There is no closed-form solution to exotic options under a regime switching model. The numerical methods to solve a system of pricing partial differential equations is complex and computational time could be substantial.

In this paper, we investigate the pricing of European-style
lookback options when the price dynamics of the underlying
risky asset are governed by a Markov-modulated Geometric Brownian motion. The Markov-modulated Geometric Brownian motion generalizes the
Geometric Brownian motion by replacing the constant market parameters with the corresponding
market parameters depending on the states of a continuous-time Markov chain model.
The Markov-modulated
model can provide a more realistic way to describe and explain the market
environment. It has been mentioned in Yao et al. (2003) that
it is of practical importance to allow the market
parameters to respond to the movements of the general market levels
since the trend of general market levels is a key factor which
governs the price movements of individual risky assets.
Markov-modulated, or regime switching, models provide one possible
way to model the situation where the market parameters depend
on a market mode which switches among a finite number
of states and reflects the state of the underlying economy,
the macro-economic condition, the general mood of the investors
in the market, business cycles and other economic factors
(See Yao et al. (2003)). By introducing the Markov-modulated
Geometric Brownian motion model, we can model the structural changes in the
volatility of the risky assets and the relationship between the
stock price and the volatility due to the change in the market regime,
in particular, economic business cycles. We derive an analytical solution for lookback options
by means of the homotopy analysis method (HAM).
HAM was initially suggested by Ortega and Rheinboldt (1970) and has been successfully used to solve a number of heat transfer problems, see Liao and Zhu (1999). Zhu (2006) proposed to adopt HAM to obtain an analytic pricing formula for American options in the Black-Scholes model. Gounden and O'Hara (2010) extended the work of Zhu to pricing an American-style Asian option of floating strike type in the Black-Scholes model. Leung (2013) used HAM to derive an analytic formula for lookback options under stochastic volatility.
The market described by the Markov-modulated Geometric Brownian motion is incomplete in
general as there is an additional source of uncertainty due to the
Markov chain. There is an infinity of equivalent
martingale measures and, hence, a range of arbitrage-free option prices.
Guo (2001) augmented the market described by a Markov-modulated
GBM by introducing a set of Arrow-Debreu's securities in order to
complete the market. Buffington and Elliott (2001a, b) followed the approach
of Guo (2001) to derive the risk-neutral dynamics of the Markov-modulated GBM.
Elliott et al. (2005) introduced the regime switching Esscher transform
to determine an equivalent martingale measure when the price dynamics of the
underlying risky asset are governed by a Markov-modulated GBM. They
justified the choice of the martingale measure by
minimizing the relative entropy. In this paper, we will adopt the regime switching Esscher transform to determine an equivalent martingale measure. We will work in this framework to price a lookback option.

This paper is organized as follows:
Section $2$ describes the dynamics of the asset price under the Markov-modulated Geometric Brownian motion.
Section $3$ describes the regime switching Esscher transform and formulates the partial differential equation system for the price of a European-style vanilla option. Section $4$ formulates a floating-strike lookback option. Section $5$ derives an exact, closed-form solution for the floating-strike lookback option. Section $6$ briefly discusses a fixed strike lookback option.  The final section draws a conclusion.
\section{Asset Price Dynamics}
Consider a complete probability space $(\Omega, {\cal F}, {\cal P})$,
where $\cal P$ is a real-world probability measure.
Let $\cal T$ denote the time index set $[0,T]$  of the model.
Write $\{W_t\}_{t \in \cal T}$ for a standard Brownian motion on
$(\Omega, {\cal F}, {\cal P})$. Suppose the states of an economy
are modelled by a finite state continuous-time Markov chain
$\{X_t\}_{t \in \cal T}$ on $(\Omega, {\cal F}, {\cal P})$. Without loss of generality, we
can identify the state space of $\{X_t\}_{t \in \cal T}$
with a finite set of unit vectors ${\cal X} := \{e_1, e_2, \dots, e_N\}$,
where $e_i = (0, \dots, 1, \dots, 0) \in \cal R^{N}$.
We suppose that $\{X_t\}_{t \in \cal T}$ and $\{W_t\}_{t \in \cal T}$
are independent.

Let $A$ be the generator $[a_{ij}]_{i,j =1, 2, \dots, N}$ of the
Markov chain process. From Elliott et al. (1994), we
have the following semi-martingale representation theorem for
$\{X_t\}_{t \in \cal T}$:
\renewcommand{\theequation}{2.\arabic{equation}}
\setcounter{equation}{0}
\begin{eqnarray}
X_t = X_0 + \int^{t}_{0} A X_s d s + M_t \ ,
\end{eqnarray}
where $\{M_t\}_{t \in \cal T}$ is an ${\cal R}^N$-valued
martingale increment process with respect to the filtration
generated by $\{X_t\}_{t \in \cal T}$.

We consider a financial model with two primary traded assets, namely
a money market account $B$ and a risky asset or stock $S$. Suppose
the market is frictionless; the borrowing and lending interest rates
are the same; the investors are price-takers.

The instantaneous market interest rate $\{r(t,
X_t)\}_{t \in \cal T}$ of the bank account is given by:
\begin{eqnarray}
r_t := r(t, X_t) = <r, X_t> \ ,
\end{eqnarray}
where $r := (r_1, r_2, \dots, r_N)$ with $r_i > 0$ for each $i= 1, 2, \dots, N$ and $<\cdot, \cdot>$ denotes the inner product in ${\cal R}^{N}$.

In this case, the dynamics of the price process $\{B_t\}_{t \in
\cal T}$ for the bank account are described by:
\begin{eqnarray}
d B_t = r_t B_t dt \ ,  \quad   B_0 = 1 \ .
\end{eqnarray}
Suppose the stock appreciation rate $\{\mu_t\}_{t \in \cal
T}$ and the volatility $\{\sigma_t\}_{t \in \cal T}$ of $S$
 depend on $\{X_t\}_{t \in \cal T}$ and are described by:
\begin{eqnarray}
\mu_t := \mu (t, X_t) = <\mu, X_t> \ , \quad \ \sigma_t := \sigma (t, X_t) = <\sigma,
X_t> \ ,
\end{eqnarray}
where $\mu := (\mu_1, \mu_2, \dots, \mu_N)$, $\sigma := (\sigma_1,
\sigma_2, \dots, \sigma_N)$ with $\sigma_i > 0$ for each $i = 1, 2,
\dots, N$ and $<\cdot, \cdot>$ denotes the inner product in ${\cal
R}^{N}$.

We assume that the price dynamics of the underlying risky asset
$S$ are governed by the Markov-modulated Geometric Brownian motion
:
\begin{eqnarray}
d S_t = \mu_t S_t d t + \sigma_t S_t d W_t \ ,
\quad
S_0 = s_{0}.
\end{eqnarray}
\section{Risk-Neutral Measure and a European-style Vanilla Option}
\renewcommand{\theequation}{3.\arabic{equation}}
\setcounter{equation}{0}
In this section, we describe the
regime-switching Esscher transform introduced in Elliott et al.
(2005) in the context of a Markovian regime-switching Black-Scholes-Merton
economy. Here we employ the pricing methodology in Elliott et al. (2005) to determine an
equivalent martingale measure.

For each $t \in {\cal T}$, let $\theta_{t}$ denote a ${\cal G} (t)$-measurable
random variable, which represents a regime-switching Esscher parameter and is
defined by:
\begin{eqnarray}
\theta_{t} := \left <\btheta, {\bf X} (t) \right > \ ,
\end{eqnarray}
where $\btheta := (\theta_1, \theta_2, \dots, \theta_{N})^{\prime} \in \Re^{N}$ and
$\theta_{i}, i=1,2,\dots, N \in (-\infty, \infty)$.

Let $(\theta \cdot W) (t) := \int^{t}_{0} \theta_{u} d W_{u}$, for each $t \in {\cal T}$.
Define a $G$-adapted density process $\Lambda^{\theta} := \{ \Lambda^{\theta} (t) | t \in {\cal T} \}$
as below:
\begin{eqnarray}
\Lambda^{\theta} (t) := \frac { e^{  (\theta \cdot W) (t) }  }
{ E [ e^{ (\theta \cdot W) } (t) | {\cal F}^{\bf X} (t) ] } \ , \quad t \in {\cal T} \ .
\end{eqnarray}
Here $E [\cdot]$ represents expectation under ${\cal P}$.

Then, applying It\^o's differentiation rule on $e^{  (\theta \cdot W) (t) }$ and conditioning on ${\cal F}^{\bf X} (t)$,
\begin{eqnarray}
E [ e^{(\theta \cdot W)} (t) | {\cal F}^{\bf X} (t) ] = \exp \bigg ( \frac {1} {2} \int^{t}_{0} \theta^2_{u} d u  \bigg ) \ .
\end{eqnarray}
So,
\begin{eqnarray}
\Lambda^{\theta} (t) := \exp \bigg ( \int^{t}_{0} \theta_{u} d W_{u} - \frac {1} {2} \int^{t}_{0} \theta^2_{u} d u \bigg )  \ , \quad t \in {\cal T} \ .
\end{eqnarray}
Since $\theta_i < \infty$, for each $i = 1, 2,\dots, N$,
\begin{eqnarray*}
E \bigg [  \exp \bigg ( \frac {1} {2} \int^{T}_{0} \theta^2_{t} d t  \bigg  )  \bigg  ] < \infty \ .
\end{eqnarray*}
So, the Novikov condition is satisfied and $\Lambda^{\theta}$ is a $(G, {\cal P})$-martingale. Hence,
\begin{eqnarray*}
E [ \Lambda^{\theta} (T) ] = 1 \ .
\end{eqnarray*}
Now, we define the regime-switching Esscher transform ${\cal Q}_{\theta} \sim \cal P$ on ${\cal G} (T)$
associated with a time-indexed family of Esscher parameters $\theta := \{ \theta_{t} | t \in {\cal T}  \}$ by setting:
\begin{eqnarray}
\frac {d {\cal Q}_{\theta}} {d {\cal P}} := \Lambda (T) \ .
\end{eqnarray}
Suppose ${\tilde \theta} :=  \{ {\tilde \theta}_{t} | t \in {\cal T} \}$ denotes a time-indexed
family of risk-neutral regime-switching Esscher parameters. Let ${\tilde S} := \{ {\tilde S}_{t} | t \in {\cal T} \}$
denote the discounted price of the risky share such that ${\tilde S}_{t} := e^{ - \int_{0}^{t}r_u du} S_{t}$, for each
$t \in {\cal T}$. Then,  the martingale condition is given by considering an enlarged filtration as follows:
\begin{eqnarray}
{\tilde S}_{u} = E^{{\tilde \theta}} [ {\tilde S}_{t}  | {\cal G} (u) ] \ , \quad  \mbox {for any $t, u \in {\cal T}$ with $t \ge u$} \ , \quad \mbox{${\cal P}$-a.s.} \ ,
\end{eqnarray}
where $E^{\tilde \theta} [\cdot]$ denotes expectation with respect to ${\cal Q}_{\tilde \theta}$.

\noindent
By setting $u = 0$,
\begin{eqnarray}
S_{0} = E^{{\tilde \theta}} [ e^{-\int_{0}^{t}r_{u}du }  S_{t}  | {\cal F}^{\bf X} (t) ] \ , \quad  \mbox {for any $t \in {\cal T}$} \ , \quad \mbox{${\cal P}$-a.s.}
\end{eqnarray}
Let $E (t) := E^{{\tilde \theta}} [ e^{-\int_{0}^{t} r_{u}du}  S_{t}  | {\cal F}^{\bf X} (t) ]$, for each $t \in {\cal T}$.  Then, $E (t)$ is an almost surely constant
random variable taking value $S_{0}$ almost surely under the measure ${\cal P}$ (i.e., ${\cal P} ( E (t) = S_{0} ) = 1$, for each $t \in {\cal T}$).
When there is no regime switching,
the condition (3.7) coincides with the one presented in Gerber and Shiu (1994).

It has been shown in Elliott et al. (2005) that the martingale condition (3.7) is satisfied if and only if ${\tilde \theta}$ satisfies
\begin{eqnarray}
{\tilde \theta}_{t} = \frac {r_t - \mu_{t}} {\sigma_{t}} = \sum^{N}_{i = 1} \bigg ( \frac {r_{t} - \mu_i} {\sigma_i} \bigg ) \left < {\bf X} (t), {\bf e}_i \right > \ , \quad t \in {\cal T} \ .
\end{eqnarray}
Since $| {\tilde \theta} (t) | < \infty$, for each $t \in [0, T]$, ${\cal P}$-a.s., the Novikov condition
\begin{eqnarray*}
E \bigg [ \exp  \bigg ( \frac {1} {2} \int^{T}_{0} {\tilde \theta}^2_{t} d t  \bigg )  \bigg  ] < \infty  \ ,
\end{eqnarray*}
is satisfied. So, $\Lambda^{\tilde \theta}$ is a $(G, {\cal P})$-martingale, and
\begin{eqnarray*}
E [ \Lambda^{\tilde \theta} (T) ] = 1  \ .
\end{eqnarray*}
Then, the risk-neutral regime-switching Esscher transform ${\cal Q}_{\tilde \theta}$ is defined by setting:
\begin{eqnarray}
\frac {d {\cal Q}_{\tilde \theta}} {d {\cal P}} &=& \Lambda^{\tilde \theta} (T) \nonumber\\
&=&  \exp \bigg [\int^{t}_{0} \bigg (\frac {r_{u} - \mu_{u} } {\sigma_{u} }
\bigg ) d W_{u} - \frac{1}{2}\int^{t}_{0} \bigg (\frac {r_{u} - \mu_{u} } {\sigma_{u}}
\bigg )^2 d u \bigg] \ .
\end{eqnarray}
Then, using Girsanov's theorem,
\begin{eqnarray*}
{\tilde W}_{t} = W_{t} - \int^{t}_{0} \bigg (\frac {r_{u} - \mu_{u}} {\sigma_{u} } \bigg )  d u \ , \quad t \in {\cal T} \ ,
\end{eqnarray*}
is a standard Brownian motion with respect to $G$ under ${\cal Q}_{{\tilde \theta}}$.

We suppose that ${\tilde W}$ and ${\bf X}$ are independent under ${\cal Q}_{{\tilde \theta}}$.
Then, the probability law of the chain ${\bf X}$ is invariant under the measure change.

The price dynamics of the underlying risky share under ${\cal Q}_{{\tilde \theta}}$ are governed by
\begin{eqnarray}
d S_{t} = r_{t} S_{t} d t + \sigma_{t} S_{t} d {\tilde W}_{t} \ .
\end{eqnarray}
Write ${\tilde {\cal G}} (t)$ for the $\sigma$-field ${\cal F}^W (t) \vee {\cal F}^{\bf X} (T)$, for each
$t \in {\cal T}$. Then, given ${\tilde {\cal G}} (t)$, a conditional price of the option $V$ is:
\begin{eqnarray}
V (t) = E^{\tilde \theta} [ e^{-\int_{t}^{T} r_{u}du}  V (S_{T}) | {\tilde {\cal G}} (t) ] \ .
\end{eqnarray}
Given $S_{t} = s$ and ${\bf X} (t) = {\bf x}$, a price of the option $V$ is:
\begin{eqnarray}
V (t, s, {\bf x}) = E^{\tilde \theta} [ e^{-\int_{t}^{T} r_{u}du}  V (S_{T})  | S_{t} = s, {\bf X} (t) = {\bf x} ] \ .
\end{eqnarray}
Let $V_i := V(t, s, {\bf e}_i)$, for each $i = 1, 2,\dots, N$. Write ${\bf V} := (V_1, V_2,\dots, V_N)^{\prime}$,
so $V(t, s, {\bf x})= \left <{\bf V}, {\bf x} \right >$.  Then, Buffington and Elliott (2002a,b)
derives the following regime-switching PDE governing the
evolution of the price of the option $V := V (t, s, {\bf x})$:
\begin{eqnarray}
- r_{t} V + \frac {\partial V} {\partial t} + r_{t} s \frac {\partial V} {\partial s} + \frac {1} {2} \sigma^2_{t}
s^2 \frac {\partial^2 V} {\partial s^2} + \left <{\bf V}, {\bf A} {\bf x} \right > = 0 \ ,
\end{eqnarray}
with terminal condition:
\begin{eqnarray}
V(T, s, {\bf x}) = V(s) \ .
\end{eqnarray}
So, if ${\bf X} (t) := {\bf e}_i$ ($i = 1, 2,\dots, N $),
\begin{eqnarray}
\mu_{t} = \mu_i \ , \quad V (t, s, {\bf x})  = V (t, s, {\bf e}_i) := V_i \ ,
\end{eqnarray}
and $V_i$ ($i = 1, 2, \dots, N$) satisfy the following system of PDEs:
\begin{eqnarray}
- r_{i} V_i + \frac {\partial V_i} {\partial t} + r_{i} s \frac {\partial V_i} {\partial s} + \frac {1} {2} \sigma^2_i s^2 \frac
{\partial^2 V_i} {\partial s^2} + \left <{\bf V}, {\bf A} {\bf e}_i \right > = 0 \ ,
\end{eqnarray}
with the terminal condition:
\begin{equation}
V(T, s, {\bf e}_i)= V(s) \ ,  \quad \ i = 1, 2,\dots, N \  .
\end{equation}
\section{Lookback Option}
\renewcommand{\theequation}{4.\arabic{equation}}
\setcounter{equation}{0}
In this section, we now turn to the pricing of lookback options in a regime switching model. We shall adopt the risk-neutral price dynamics
of the risky stock under ${\cal Q}_{{\tilde \theta}}$ specified  in the last section to evaluate the price of lookback
 options. In particular, we consider a floating strike lookback option under a regime switching model. The payoff of this option is the difference between the maximum asset price over the time between initiation and expiration and the asset price at expiration. The maximum of the asset price up to time $t$ is denoted by
\begin{eqnarray}
Y(t)=\max_{0\le u\le t} S_{u}.
\end{eqnarray}
Then the payoff of the lookback option at expiration time $T$ is
\begin{eqnarray}
V(T)=Y(T)-S_{T}.
\end{eqnarray}
Given $S_{t} = s$, $Y(t)=y$ and ${\bf X} (t) = {\bf x}$, a price of the lookback option $V$ is:
\begin{eqnarray}
V (t, s,y,{\bf x}) = E^{\tilde \theta} [ e^{-\int_{t}^{T} r_{u}du}(Y(T)-S_{T})  | S_{t} = s, Y(t)=y, {\bf X} (t) = {\bf x} ] \ .
\end{eqnarray}
Applying the Feynman-Kac formula to the above equation, then $V (t, s,y,{\bf x})$ satisfies the system of partial differential equations (PDEs)
\begin{eqnarray}
\frac{\partial V}{\partial t}+r_{t}s\frac{\partial V}{\partial s}+\frac{1}{2}\sigma_{t}^{2}s^{2} \frac{\partial^{2}V}{\partial s^{2}}-r_{t}V+ \left <{\bf V}, {\bf A} {\bf x} \right > = 0 \ ,
\end{eqnarray}
in the region $\{(t,s,y); 0\le t< T, 0\le s \le y \}$ and satisfies the boundary conditions
\begin{eqnarray}
V(T,s,y,{\bf x})=f(s,y)=y-s, \quad 0\leq s\leq y \,
\end{eqnarray}
\begin{eqnarray}
\frac{\partial V}{\partial y}(t,y,y,{\bf x}) =0, \quad 0\leq t\le T,\quad y>0\,
\end{eqnarray}
Consequently, if ${\bf X} (t) := {\bf e}_i$ ($i = 1, 2,\dots, N $),
\begin{eqnarray}
\mu_{t} = \mu_i \ , \quad V (t,s,y,{\bf x})  = V (t,s,y,{\bf e}_i) := V_i \ ,
\end{eqnarray}
and $V_i$ ($i = 1, 2, \dots, N$) satisfy the following system of PDEs:
\begin{eqnarray}
- r_{i} V_i + \frac {\partial V_i} {\partial t} + r_{i} s \frac {\partial V_i} {\partial s} + \frac {1} {2} \sigma^2_i s^2 \frac
{\partial^2 V_i} {\partial s^2} + \left <{\bf V}, {\bf A} {\bf e}_i \right > = 0 \ ,
\end{eqnarray}
with the boundary conditions:
\begin{eqnarray}
V(T,s,y,{\bf e}_i )=f(s,y)=y-s,  \quad 0\leq s\leq y, \quad \ i = 1, 2,\dots, N \,
\end{eqnarray}
\begin{eqnarray}
\frac{\partial V}{\partial y}(t,y,y,{\bf e}_i) =0, \quad 0\leq t\le T,\quad y>0, \quad \ i = 1, 2,\dots, N.
\end{eqnarray}
\section{A closed-form formula}
\renewcommand{\theequation}{5.\arabic{equation}}
In this section, we restrict ourselves to a special case with the number of regimes $N$ being $2$ in order
to simplify our discussion. By means of the homotopy analysis method, we derive a closed-form solution for a floating strike lookback option under a regime switching model.
The payoff of the floating strike lookback option has a linear homogeneous property:
\begin{eqnarray}
f(s,y)=s g\bigg(\ln\big(\frac{y}{s}\big) \bigg)\
\end{eqnarray}
where
\begin{eqnarray}
g(z)=e^{z}-1, \quad z=\ln\big(\frac{y}{s}\big).
\end{eqnarray}
This linear homogeneous property along with the transformation $z=\ln(\frac{y}{s})$ and $U_{i}=\frac{V_{i}}{s}, i=1,2$ transform the system of equations (4.8)-(4.10) into
\begin{eqnarray}
\left\{\begin{array}{lll}
{\cal L}_{1} U_{1}(t,z)= a_{11}\big(U_{2}(t,z)-U_{1}(t,z)\big),\quad 0\leq t\leq T,\quad z>0
 \\
 U_{1}(T,z)=g(z)
 \\
\frac{\partial U_{1}}{\partial z}(t,z)|_{z=0}=0
\end{array} \right.
\end{eqnarray}
where
\begin{eqnarray}
 {\cal L}_{1}=\frac{\partial}{\partial t}+\frac{1}{2}\sigma_{1}^{2}\frac{\partial^{2}}{\partial z^{2}}-(r_{1}+\frac{\sigma_{1}^{2}}{2})\frac{\partial }{\partial z} \,
\end{eqnarray}
and
\begin{eqnarray}
\left\{\begin{array}{lll}
 {\cal L}_{2} U_{2}(t,z)=a_{22}\big(U_{1}(t,z)-U_{2}(t,z)\big),\quad 0\leq t\leq T,\quad z>0
 \\
 U_{2}(T,z)=g(z)
 \\
\frac{\partial U_{2}}{\partial z}(t,z)|_{z=0}=0
\end{array} \right.
\end{eqnarray}
where
\begin{eqnarray}
 {\cal L}_{2}=\frac{\partial}{\partial t}+\frac{1}{2}\sigma_{2}^{2}\frac{\partial^{2}}{\partial z^{2}}-(r_{2}+\frac{\sigma_{2}^{2}}{2})\frac{\partial }{\partial z}.
\end{eqnarray}
Following the same line as Leung (2013), the homotopy analysis method is adopted to solve $ U_{i}(t,z), i=1,2$ from equations (5.13) and (5.15).

Now we introduce an embedding parameter $p\in [0, 1]$ and construct unknown functions
${\bar{U}}_{i}(t,z,p), i=1, 2$ that satisfy the following differential systems:
\begin{eqnarray}
\left\{\begin{array}{lll}
(1-p) {\cal L}_{1}[{\bar{U}}_{1}(t,z,p)-{\bar{U}}_{1}^{0}(t,z)]=-p\bigg\{{\cal A}_{1}[{\bar{U}}_{1}(t,z,p), {\bar{U}}_{2}(t,z,p)]  \bigg\}
 \\
 {\bar{U}}_{1}(t, z, p)=g(z)
 \\
\frac {\partial {\bar{U}}_{1}} {\partial z} (t, 0, p)= (1-p) \frac {\partial {\bar{U}}_{1}^{0}} {\partial z} (t,0)
\end{array} \right.
\end{eqnarray}
\begin{eqnarray}
\left\{\begin{array}{lll}
(1-p) {\cal L}_{2}[{\bar{U}}_{2}(t,z,p)-{\bar{U}}_{2}^{0}(t,z)]=-p\bigg\{{\cal A}_{2}[{\bar{U}}_{1}(t,z,p), {\bar{U}}_{2}(t,z,p)]  \bigg\} \,
 \\
 {\bar{U}}_{2}(t, z, p)=g(z) \,
 \\
\frac {\partial {\bar{U}}_{2}} {\partial z} (t, 0, p)= (1-p) \frac {\partial {\bar{U}}_{2}^{0}} {\partial z} (t,0) \,
\end{array} \right.
\end{eqnarray}
Here ${\cal L}_{i}, i=1,2$ is a differential operator defined as
\begin{eqnarray}
{\cal L}_{i}=\frac{\partial}{\partial t}+\frac{1}{2}\sigma_{i}^{2}\frac{\partial^{2}}{\partial z^{2}}-(r_{i}+\frac{\sigma_{i}^{2}}{2})\frac{\partial }{\partial z} \,
\end{eqnarray}
and ${\cal A}_{i}, i=1,2$ are functionals defined as
\begin{eqnarray}
{\cal A}_{1}[{\bar{U}}_{1}(t,z,p),{\bar{U}}_{2}(t,z,p)]={\cal L}_{1}({\bar{U}}_{1})-a_{11}({\bar{U}}_{1}-{\bar{U}}_{2})
\end{eqnarray}
\begin{eqnarray}
{\cal A}_{2}[{\bar{U}}_{1}(t,z,p),{\bar{U}}_{2}(t,z,p)]={\cal L}_{2}({\bar{U}}_{2})-a_{22} ({\bar{U}}_{2}-{\bar{U}}_{1})
\end{eqnarray}
With $p=1$, we have
\begin{eqnarray}
\left\{\begin{array}{lll}
{\cal L}_{1}({\bar{U}}_{1})=a_{11}({\bar{U}}_{1}-{\bar{U}}_{2}) \,
 \\
 {\bar{U}}_{1}(t, z, 1)=g(z) \,
 \\
\frac {\partial {\bar{U}}_{1}} {\partial z} (t, z, 1)|_{z=0}= 0 \,
\end{array} \right.
\end{eqnarray}
\begin{eqnarray}
\left\{\begin{array}{lll}
{\cal L}_{2}({\bar{U}}_{2})=a_{22} ({\bar{U}}_{2}-{\bar{U}}_{1}) \,
 \\
 {\bar{U}}_{2}(t, z, 1)=g(z) \,
 \\
\frac {\partial {\bar{U}}_{2}} {\partial z} (t, z, 1)|_{z=0}= 0 \,
\end{array} \right.
\end{eqnarray}
Comparing with (5.13) and (5.15), it is obvious that ${\bar{U}}_{i}(t,z,1), i=1,2$ are equal to our searched solutions $U_{i}(t,z), i=1,2$.

Now we set $p=0$, the equations (5.17) and (5.18) become
\begin{eqnarray}
\left\{\begin{array}{lll}
 {\cal L}_{1}[{\bar{U}}_{1}(t,z,p)]={\cal L}_{1}[{\bar{U}}_{1}^{0}(t,z)] \,
 \\
 {\bar{U}}_{1}(t, z, 0)=g(z) \,
 \\
\frac {\partial {\bar{U}}_{1}} {\partial z} (t, 0, 0)= \frac {\partial {\bar{U}}_{1}^{0}} {\partial z} (t,0) \,
\end{array} \right.
\end{eqnarray}
\begin{eqnarray}
\left\{\begin{array}{lll}
 {\cal L}_{2}[{\bar{U}}_{2}(t,z,p)]={\cal L}_{2}[{\bar{U}}_{2}^{0}(t,z)] \,
 \\
 {\bar{U}}_{2}(t, z, 0)=g(z) \,
 \\
\frac {\partial {\bar{U}}_{2}} {\partial z} (t, 0, 0)=  \frac {\partial {\bar{U}}_{2}^{0}} {\partial z} (t,0) \,
\end{array} \right.
\end{eqnarray}
${\bar{U}}_{i}(t,z,0), i=1,2$ will be equal to ${\bar{U}}_{i}^{0}(t,z)$ when ${\bar{U}}_{i}^{0}(T,z)=g(z), i=1,2$. ${\bar{U}}_{i}^{0}(t,z)$ is known as the initial guess of $U_{i}(t,z)$.
Following the same line as Leung (2013), ${\bar{U}}_{i}^{0}(t,z)$ is chosen as the solution of the following PDEs:
\begin{eqnarray}
\left\{\begin{array}{lll}
 {\cal L}_{1}[{\bar{U}}_{1}^{0}(t,z)]=0 \,
 \\
 {\bar{U}}_{1}^{0}(0, z)=g(z) \,
 \\
\frac {\partial {\bar{U}}_{1}^{0}} {\partial z} (t, z)|_{z=0}= 0 \,
\end{array} \right.
\end{eqnarray}
\begin{eqnarray}
\left\{\begin{array}{lll}
 {\cal L}_{2}[{\bar{U}}_{2}^{0}(t,z)]=0 \,
 \\
 {\bar{U}}_{2}^{0}(0, z)=g(z) \,
 \\
\frac {\partial {\bar{U}}_{2}^{0}} {\partial z} (t,z)|_{z=0}=0 \,
\end{array} \right.
\end{eqnarray}
Note that $s{\bar{U}}_{1}^{0}(t,z)$ is the price of the floating strike lookback put option under the Black-Scholes-Merton model. Its explicit, closed-form formula is given in Goldman, et al. (1979):
\begin{eqnarray}
 {\bar{U}}_{i}^{0}(t,z)=e^{z}e^{-r_{i}(T-t)}N(-d_{M}^{-})-N(-d_{M}^{+})+\frac{{\sigma}_{i}^2}{2r_{i}}\big[N(d_{M}^{+})-e^{-r_{i} (T-t)}e^{\frac{2r_{i}z}{{\sigma}_{i}^2}}N(d_{M}^{\prime})\big],
\end{eqnarray}
where
\begin{eqnarray}
d_{M}^{\pm}=\frac{-z+(r_{i}\pm\frac{{\sigma}_{i}^2}{2}(T-t)}{\sqrt{{\sigma}_{i}^2(T-t)}},
\end{eqnarray}
\begin{eqnarray}
d_{M}^{\prime}=d_{M}^{+}-\frac{2r_{i}}{{\sigma}_{i}}\sqrt{T-t}
\end{eqnarray}
and $N(.)$ is the cumulative standard normal distribution
\begin{eqnarray}
N(y)=\frac{1}{2\pi}\int_{-\infty}^{y}e^{-\frac{z^2}{2}}dz.
\end{eqnarray}
To find the values of ${\bar{U}}_{i}(t, z, 1), i=1,2$,
 we can expand the functions ${\bar{U}}_{i}(t, z, p)$ as a Taylor's series expansion of $p$
 \begin{eqnarray}
\left\{\begin{array}{lll}
{\bar{U}}_{1}(t, z, p)=\sum_{m=0}^{\infty}\frac{{\bar{U}}_{1}^{m}(t, z)}{m !}p^{m}\,
\\
{\bar{U}}_{2}(t, z, p)=\sum_{m=0}^{\infty}\frac{{\bar{U}}_{2}^{m}(t, z)}{m !}p^{m}\,
\end{array} \right.
\end{eqnarray}
 where
 \begin{eqnarray}
\left\{\begin{array}{lll}
{\bar{U}}_{1}^{m}(t, z)=\frac{\partial ^{m}}{\partial p^{m}} {\bar{U}}_{1}(t, z, p)|_{p=0}\,
\\
{\bar{U}}_{2}^{m}(t, z)=\frac{\partial ^{m}}{\partial p^{m}} {\bar{U}}_{2}(t, z, p)|_{p=0}\,
\end{array} \right.
\end{eqnarray}
To find ${\bar{U}}_{1}^{m}(t, z)$ and ${\bar{U}}_{2}^{m}(t, z)$ in equation (5.32), we put (5.32) into (5.17) and (5.18) respectively and obtain the following recursive relations:
\begin{eqnarray}
\left\{\begin{array}{lll}
{\cal L}_{1}({\bar{U}}_{1}^{m})=a_{11}({\bar{U}}_{1}^{m-1}-{\bar{U}}_{2}^{m-1})\quad m=1,2,..., \,
 \\
 {\bar{U}}_{1}^{m}(T, z)=0 \,
 \\
\frac {\partial {\bar{U}}_{1}^{m}} {\partial z} (t, z)|_{z=0}= 0 \,
\end{array} \right.
\end{eqnarray}
\begin{eqnarray}
\left\{\begin{array}{lll}
{\cal L}_{2}({\bar{U}}_{2}^{m})=a_{22} ({\bar{U}}_{2}^{m-1}-{\bar{U}}_{1}^{m-1})\quad m=1,2,..., \,
 \\
 {\bar{U}}_{2}^{m}(T, z)=0 \,
 \\
\frac {\partial {\bar{U}}_{2}^{m}} {\partial z} (t, z)|_{z=0}= 0 \,
\end{array} \right.
\end{eqnarray}
We introduce the following transformations:
\begin{eqnarray}
\tau=T-t,\quad \alpha_{i}=\frac{2 r_{i}}{\sigma_{i}^2},
\end{eqnarray}
and
\begin{eqnarray}
{\bar{U}}_{i}^{m}(t, z)=e^{[-\frac{1}{8}\sigma_{i}^2(\alpha_{i}+1)^2 \tau + \frac{1}{2}(\alpha_{i}+1)z]}{\hat{U}}_{i}^{m}(t, z).
\end{eqnarray}
 We can rewrite equations (5.34) and (5.35) in the form of standard nonhomogeneous diffusion equations
 \begin{eqnarray}
\left\{\begin{array}{lll}
\frac{\partial {\hat{U}}_{1}^{m}}{\partial \tau}-\frac{1}{2}\sigma_{1}^2\frac{\partial^{2} {\hat{U}}_{1}^{m}}{\partial z^{2}}=a_{11} e^{[-\frac{1}{8}\sigma_{1}^2(\alpha_{1}+1)^2 \tau + \frac{1}{2}(\alpha_{1}+1)z]}\big({\bar{U}}_{1}^{m-1}-{\bar{U}}_{2}^{m-1}\big) \,
 \\
 {\hat{U}}_{1}^{m}(0, z)=0 \,
 \\
\frac {\partial {\hat{U}}_{1}^{m}} {\partial z} (\tau, 0)+\frac{1}{2}(\alpha_{1}+1){\hat{U}}_{1}^{m}(\tau, 0)= 0 \,
\end{array} \right.
\end{eqnarray}
\begin{eqnarray}
\left\{\begin{array}{lll}
\frac{\partial {\hat{U}}_{2}^{m}}{\partial \tau}-\frac{1}{2}\sigma_{2}^2\frac{\partial^{2} {\hat{U}}_{2}^{m}}{\partial z^{2}}=a_{22} e^{[-\frac{1}{8}\sigma_{2}^2(\alpha_{2}+1)^2 \tau + \frac{1}{2}(\alpha_{2}+1)z]}\big({\bar{U}}_{2}^{m-1}-{\bar{U}}_{1}^{m-1}\big) \,
 \\
 {\hat{U}}_{2}^{m}(0, z)=0 \,
 \\
\frac {\partial {\hat{U}}_{2}^{m}} {\partial z} (\tau, 0)+\frac{1}{2}(\alpha_{2}+1){\hat{U}}_{2}^{m}(\tau, 0)= 0 \,
\end{array} \right.
\end{eqnarray}
The system of PDEs (5.38) and (5.39) has a well-known closed-form solution respectively:
\begin{eqnarray}
 {\hat{U}}_{1}^{m}(\tau, z)&=& a_{11}\int_{0}^{\tau}\int_{0}^{\infty} \exp\big\{-\frac{1}{8}\sigma_{1}^2(\alpha_{1}+1)^2 u + \frac{1}{2}(\alpha_{1}+1)\xi\big\}\nonumber\\
 &\times& \big({\bar{U}}_{1}^{m-1}(T-u,\xi)-{\bar{U}}_{2}^{m-1}(T-u,\xi)\big)G_{1}(\tau-u,z,\xi)d\xi du,
\end{eqnarray}
\begin{eqnarray}
 {\hat{U}}_{2}^{m}(\tau, z)&=&a_{22}\int_{0}^{\tau}\int_{0}^{\infty} \exp\big\{-\frac{1}{8}\sigma_{2}^2(\alpha_{2}+1)^2 u + \frac{1}{2}(\alpha_{2}+1)\xi \big\}\nonumber\\
 &\times& \big({\bar{U}}_{2}^{m-1}(T-u,\xi)-{\bar{U}}_{1}^{m-1}(T-u,\xi)\big)G_{2}(\tau-u,z,\xi)d\xi du,
\end{eqnarray}
where
\begin{eqnarray}
G_{1}(t,z,\xi)&=&\frac{1}{\sqrt{2\pi\sigma_{1}^2 t}}\bigg\{\exp\bigg[\frac{-(z-\xi)^{2}}{2\sigma_{1}^2 t} \bigg]+\exp\bigg[\frac{-(z+\xi)^{2}}{2\sigma_{1}^2 t} \bigg]\nonumber\\
&+&2\kappa_{1}\int_{0}^{\infty}\exp\bigg[-\frac{(z+\xi+\eta_{1})^2}{2\sigma_{1}^2 t}+\kappa_{1}\eta_{1}\bigg]d\eta_{1}\bigg\}\nonumber\\
&=&\frac{1}{\sqrt{2\pi\sigma_{1}^2 t}}\bigg\{\exp\bigg[\frac{-(z-\xi)^{2}}{2\sigma_{1}^2 t} \bigg]+\exp\bigg[\frac{-(z+\xi)^{2}}{2\sigma_{1}^2 t} \bigg]\nonumber\\
&+&2\kappa_{1}\exp(\kappa_{1}(-z-\xi+\frac{\sigma_{1}^{2}\kappa_{1}}{2}))N\big(\frac{\sigma_{1}^{2}\kappa_{1}t-z-\xi}{\sigma_{1}\sqrt{t}}\big)\bigg\},
\end{eqnarray}
\begin{eqnarray}
G_{2}(t, z,\xi)&=&\frac{1}{\sqrt{2\pi\sigma_{2}^2 t}}\bigg\{\exp\bigg[\frac{-(z-\xi)^{2}}{2\sigma_{2}^2 t} \bigg]+\exp\bigg[\frac{-(z+\xi)^{2}}{2\sigma_{2}^2 t} \bigg]\nonumber\\
&+&2\kappa_{2}\int_{0}^{\infty}\exp\bigg[-\frac{(z+\xi+\eta_{2})^2}{2\sigma_{2}^2 t}+\kappa_{2}\eta_{2}\bigg]d\eta_{2}\bigg\}\nonumber\\
&=&\frac{1}{\sqrt{2\pi\sigma_{2}^2 t}}\bigg\{\exp\bigg[\frac{-(z-\xi)^{2}}{2\sigma_{2}^2 t} \bigg]+\exp\bigg[\frac{-(z+\xi)^{2}}{2\sigma_{2}^2 t} \bigg]\nonumber\\
&+&2\kappa_{2}\exp(\kappa_{2}(-z-\xi+\frac{\sigma_{2}^{2}\kappa_{2}}{2}))N\big(\frac{\sigma_{2}^{2}\kappa_{2}t-z-\xi}{\sigma_{2}\sqrt{t}}\big)\bigg\},
\end{eqnarray}
and $$\kappa_{i}=\frac{1}{2}(\alpha_{i}+1), \quad \eta_{i}=e^{[-\frac{1}{8}\sigma_{i}^2(\alpha_{i}+1)^2 \tau + \frac{1}{2}(\alpha_{i}+1)z]},\quad i=1,2$$
\section{Fixed strike lookback options}
\renewcommand{\theequation}{6.\arabic{equation}}
The payoff of a fixed strike lookback option does not have a linear homogeneous property. Consequently, the dimension of the corresponding PDEs cannot be reduced.
However, a model independent put-call parity for lookback options, proposed by Wong and Kwok (2003) may be used to price a fixed strike lookback option.
Denote a fixed strike lookback call by $C_{fix}(t,s,y,K,{\bf x})$, then the put-call parity is given by
\begin{eqnarray}
C_{fix}(t,s,y,K,{\bf x})=V (t, s,y,{\bf x})+Ke^{-r_{t}(T-t)}.
\end{eqnarray}
\section{Conclusion}
We consider the pricing of the floating strike lookback option in a two-state regime switching model.
The closed-form analytical pricing formulas for the floating strike lookback option is derived by the means of the homotopy analysis method.

\end{document}